\documentclass[11pt]{article}

\usepackage[T1]{fontenc}
\usepackage[margin=1in]{geometry}
\usepackage{amsmath,amssymb}
\usepackage{graphicx}
\usepackage{authblk}
\usepackage{xcolor}
\usepackage{tikz}
\usetikzlibrary{positioning,arrows.meta,fit,backgrounds}
\usepackage{natbib}
\usepackage{hyperref}

\title{Towards foundation models for insurance risk modelling}

\author{Christopher Blier-Wong}
\affil{Department of Statistical Sciences, University of Toronto, Canada}

\date{\today}

\begin{document}

\maketitle

\begin{abstract}
Claim narratives, images and sensor data contain information about
insured risks that is difficult to use through existing actuarial models.
Foundation models learn patterns from large datasets before being adapted to
particular tasks. By turning these high-dimensional sources into variables or
numerical representations, they could help insurers use more of the information
they already collect, potentially reducing the experience needed to develop each application.
For example, a language model could identify a worsening injury in a new claim
note, allowing a reserving model to recognise the change in expected cost before
the payments reveal the deterioration. In this paper, we review language, vision,
geospatial, time series, tabular and scientific models, explaining existing
insurance applications and potential future uses. Scientific models extend
this approach to future weather and climate conditions: their simulations can
inform loss estimates once local hazards are linked to asset damage, repair
costs and insurance coverage. We propose a process to connect these model
outputs to actuarial calculations and to assess their predictive contribution,
stability and compliance with rules on information use. Evaluating these applications is difficult when final claim costs become
known only after long delays, large losses are rare or patterns
learned elsewhere fail to transfer to the target portfolio. Richer data
can reveal private information and support finer risk classification,
which can change access to insurance. Reusing the same models across
insurers also creates dependence on shared predictions and providers.

\end{abstract}


\section{Introduction}
\label{sec:introduction}

Insurers collect detailed information about policyholders, insured assets and
claims. Textual claim descriptions, images, reports and sensor data describe
risks in ways that structured actuarial variables may not capture. A claim
description can explain why treatment is continuing, for example, while a
payment record shows only the amounts paid so far. Insurers already have
access to much of this information, but often lack the modelling capability
to use its contents in loss prediction.
The information needed to estimate insured costs changes over the life of a
contract. At underwriting, insurers set premiums, coverage and deductibles
before knowing whether a claim will occur or how large it will be. As claims develop through treatment, payments and settlement, insurers
update reserve estimates using the evidence that arrives. In
capital management, they assess how losses on existing and proposed contracts
could occur together. Rare large claims, delayed outcomes and changes in risk
make these tasks difficult. Making fuller use of the available data could
help insurers recognise changes in expected costs earlier and represent
risks more accurately in their models.

Text and image models trained for specific tasks can process these sources,
but developing a separate model for each source and task is costly.
Foundation models offer a way to transfer knowledge between tasks. When a
model is trained on broad datasets, it can learn general patterns in language,
images or other data that are useful in several applications. This initial
training, called pretraining, may allow an insurer to develop a new
application with less claims experience than would be needed to train the
whole model from the beginning. Following
\citet{bommasani2021opportunities}, we use \emph{foundation model} for a model
trained at scale on broad data and reused across tasks, including models
pretrained on distributions of simulated prediction tasks.
Consider a bodily injury claim. The claim file may describe treatment and
litigation in detail while the reserving model receives numerical summaries
of claim status and a series of payment amounts. A claims adjuster may record
a worsening injury months before the payments increase. A pretrained language
model could identify that change or encode the updated claim status as
numerical inputs, allowing the reserving model to respond earlier if the
change is not already captured by the adjuster's recorded case estimate. Images
of a roof and measurements from a water sensor offer similar ways to observe
deterioration or a leak between inspections. For losses caused by future
floods or storms, physical models can instead generate hazard scenarios,
which an insurer can translate into damage and contractual payments.

A variable such as roof age can enter a pricing model
directly. An embedding, which is a numerical representation of an image or
text, can supply several inputs describing patterns that are difficult to
name individually. To understand a claim that is still developing, an adjuster can use a
retrieval model to search a database for earlier claims with similar
circumstances. Simulated
weather scenarios supply another kind of input, from which an insurer can
estimate damage and payments. These uses call for different ways to connect
the output to an actuarial calculation.
We propose a process in which the insurer selects and prepares the output,
applies the rules governing its use, and passes the resulting information to
the actuarial model. To trace changes in estimates to changes in inputs, the insurer should retain the
source, model version and validation results. To compare foundation models directly, the insurer can keep the fitted actuarial
model fixed only when their outputs have the same meaning and compatible format.

We examine existing insurance applications and potential uses of foundation
models, explaining how insurers could connect their outputs to pricing,
reserving, prevention and capital calculations. The proposed process
combines assessment of predictive performance and stability with the
rules governing which information an insurer may use.
The use of learned representations builds on our earlier work in
\citet{blierwong2021rethinking}. Insurance data are costly to collect, and
individual losses can vary widely even among similar risks. With few claims available, fitting a model with too many parameters risks
learning accidental patterns in the sample, a problem known as overfitting. We proposed first learning general
patterns from larger datasets, then transferring the resulting
representations to insurance tasks. A representation learned from images,
for example, can describe a property's surroundings before claims data are
used to estimate how those surroundings relate to losses. This separates
the data needed to learn the representation from the claims needed to fit
the actuarial model, allowing more flexible relationships in P\&C ratemaking
than the claims alone might support. Foundation models extend these principles of representation
learning and transfer through pretraining on much larger and more varied
datasets \citep{bommasani2021opportunities}. The resulting models can process
several types of insurance data without requiring an insurer to develop
every representation from its own claims. We extend the discussion beyond
ratemaking to reserving, prevention and risk assessment, and to the
consequences for the insurance industry.

Many existing foundation models could support these insurance applications. Language
and vision models handle text and images; geospatial, time series and tabular
models process location data, sequential measurements and structured data.
Case studies using transformer models already support the predictive use of
claim text \citep{troxler2022actuarial}. In reserving, such text could reveal a
new complication or the prospect of litigation before either changes the
payment pattern, allowing the reserving model to update expected remaining
payments sooner. Tabular foundation models have achieved strong results on
general prediction tasks \citep{hollmann2025accurate}, although recent
insurance comparisons do not establish a consistent advantage over boosted
trees \citep{deprez2026tabpfn}. Scientific models offer a different opportunity.
To quantify the cost of weather events in a changing climate, insurers need
forecasts of physical conditions as well as historical claims.
Weather forecasts \citep{bodnar2025foundation} can support this calculation
once the predicted conditions are translated into damage to insured assets
and payments after coverage, deductibles and limits are applied.

Section \ref{sec:landscape} explains applications to insurance data, and
Section \ref{sec:scientific} follows physical simulations through to insured
losses. Section \ref{sec:architecture} describes how pretrained models can be
combined with insurer experience. Section \ref{sec:governance} brings together
validation and governance, before Section \ref{sec:economics} examines effects
on policyholders and markets. Section \ref{sec:agenda} develops the idea of
actuarial foundation models that can be reused across insurance tasks.
Section \ref{sec:conclusion} concludes.

\section{Foundation models for insurance data}
\label{sec:landscape}
\label{sec:modalities}

An insurer may hold years of claim descriptions, images and measurements
without having a practical way to include their contents in a predictive
model. Foundation models can convert these sources into variables or
embeddings that an actuary can relate to losses. The conversion differs by
data type: a language model represents the meaning of a claim description,
whereas a vision model represents patterns in an image. In both cases, the
actuarial objective is to use the additional information to estimate insured
costs or identify opportunities to prevent damage. The applications below
explain what the models produce and how those outputs could contribute to
insurance calculations.

\subsection{Models and outputs}

Insurers already fit gradient-boosted trees and neural networks to
portfolio data for pricing, mortality and reserving
\citep{richman2021ai,richman2021aib,blierwong2021machine,wuthrich2023statistical,holvoet2025benchmark}.
Hybrid actuarial models combine a classical model with a neural network.
Combined Actuarial Neural Networks retain an actuarial predictor within the
network \citep{wuthrich2019cann,schelldorfer2019nesting}, while LocalGLMnet
allows regression coefficients to depend on the covariates
\citep{richman2023localglmnet}. These approaches preserve a connection to
familiar actuarial models while allowing more flexible relationships.
Foundation models can supply inputs to such actuarial models. Our
definition includes conversational systems, encoders that turn data into
numerical vectors, and generators of scenarios. These systems differ in
the data they process and how they are built, as well as in their outputs.
Their different outputs
require different ways of connecting them to the actuarial calculation.

Representation learning offers another way to connect neural networks and
actuarial models. An encoder learns numerical inputs that a pricing model can
use in place of, or alongside, conventional rating variables. Applications and
discussions include \citet{richman2021ai}, \citet{blierwong2021rethinking},
\citet{delong2023autoencoders} and \citet{shi2023nonlife}.
Pretraining foundation models extends this approach by learning from data
beyond the portfolio being modelled. A model may learn to recognise the
meaning of a medical description from general text, for example, while the
insurer's claims data are used to estimate how that information relates
to payments.

The outputs of foundation models that can be used as inputs to actuarial
models include
interpretable variables, embeddings, retrieved documents and simulated
scenarios. An \emph{interpretable variable} names an attribute of the source,
such as the roof material visible in an image or the injury described in a
claim note. An insurer should specify whether an \emph{injury} variable represents
a reported symptom or a confirmed diagnosis. An \emph{embedding} is a dense numerical
vector produced by an encoder from high-dimensional or unstructured data.
The encoder aims to retain useful information while reducing redundancy.
The geometry of the embedding space can place observations with similar characteristics close together. Text descriptions, for example, can be close in that space when they have similar meanings, even if they use different words. An individual coordinate may have no meaning that one can name. The
actuarial model instead uses the coordinates together as inputs for a
task such as predicting losses.

An adjuster
handling a disputed water-damage claim could search earlier claim files for
similar causes of loss or search the policy wording for the applicable
exclusion. The selected documents then help the adjuster interpret the
current claim. \emph{Simulated scenarios} describe possible states or paths
of a physical system, such as rainfall over a catchment and the resulting
water levels in a floodplain. An insurer can translate those scenarios into
losses on the buildings it covers. To estimate loss probabilities, an insurer can use scenarios sampled from a
predictive distribution of physical conditions. To examine potential
losses without assigning a
probability to an event, it can instead simulate unusually heavy rainfall.

\subsection{Language models}
\label{sec:language}

Generative language models include the GPT models used in ChatGPT and the
conversational models in the Gemini, Claude, Qwen and DeepSeek families.
They generate text and can answer questions or fill fields from a claim
description, using instructions and examples supplied with the request
\citep{brown2020language}. Embedding models instead turn text into numerical
vectors for search or prediction. Examples include Gemini Embedding
\citep{lee2025gemini} and Qwen3 Embedding \citep{zhang2025qwen}; BERT also
produces contextual text representations \citep{devlin2019bert}. An encoder
represents a claim description by a vector whose values are used together.
When reserving for bodily injury claims, that vector could retain information about treatment or
recovery that is absent from the injury type and payment amounts recorded in
structured fields.

Text generation can produce different answers to the same request. Generated
text can also be grammatically convincing while inventing facts absent from
the source, a problem known as \emph{hallucination} \citep{ji2023survey}.
With identical text, a fixed encoder and preprocessing, and deterministic
numerical execution, an insurer can reproduce an embedding without sampling
an answer. A reproducible embedding can still misrepresent the text without
an explicit assertion to check, while rewording or changes in hardware and
batching can affect the outputs (Section \ref{sec:stability}).

Insurance documents include adjuster notes, policy wordings, underwriting
applications, legal correspondence, medical descriptions and reports, such
as inspections of damaged buildings. Their rich descriptions can explain
circumstances omitted from claim codes and payment records.
\citet{lee2020actuarial} and \citet{troxler2022actuarial} examine the use of
textual descriptions alongside structured claim information. Troxler and Schelldorfer evaluate transformer models that classify
multilingual descriptions of accidents by the number of vehicles involved
or the presence of bodily injury. They also classify short descriptions
of property claims by peril. In reserving, a language model could supply updated information
as a claim develops, so that a new treatment plan or a dispute about liability
changes the estimated remaining payments.

An insurer may want to retain familiar actuarial variables while using
details that currently appear only in text. A language model can fill fields
such as injury type, treatment status and litigation stage from a claim
description. To give the same wording a consistent interpretation across claims, an
insurer should first define the fields. For example, a claimant's
description of pain, a medical diagnosis and an adjuster's assessment of the
injury support different variables. Recording all three as a confirmed diagnosis would hide whether a clinician had verified the injury, preventing the model from using that distinction in the loss estimate.
The same approach could extract variables such as roof age, construction material and occupancy from an underwriting application or inspection report.
A pricing model could then use these variables alongside the information
already collected on the application form.

A text encoder can retain information that is difficult to capture in a
short list of fields. The aim is to represent relationships in meaning through the relative
positions of text embeddings. Such a representation can supply inputs
for several tasks, rather than being chosen only for one prediction task.
Actuarial studies have used transformer representations and interpretable
text-mining methods to extract predictive information
\citep{lee2020actuarial,zappa2021text,troxler2022actuarial,xu2022bert}.
To assess whether an embedding is useful, researchers can compare
predictions made with and without it for the intended actuarial task,
such as predicting claim amounts from warranty descriptions.
Similarly, \citet{blierwong2026semantic} use a pretrained language model to
embed descriptions of structured policyholder variables, then fit a Poisson
generalised linear model (GLM) to the embeddings. They report gains over the conventional GLM, especially when training
samples are small. The model makes existing policyholder information more
useful to the pricing calculation without adding new policyholder fields.

An adjuster facing an unfamiliar claim may benefit from seeing how similar
claims developed and were resolved. The adjuster can search a database of previous claims for comparable injuries or causes of damage. A search based only on keywords can miss those examples: two claimants may describe similar injuries or causes of
damage in different words. Embeddings allow semantic retrieval of claim
documents by comparing the meanings of the descriptions. The adjuster can
then examine earlier claims with similar initial circumstances, including
the treatment, litigation or repairs that followed. Policy wording can be
searched in the same way to find clauses relevant to the present facts.
In retrieval-augmented generation, a language model uses the selected
documents to produce a response \citep{lewis2020retrieval}. The retrieved
documents provide evidence for that response and examples to inform judgment;
differences in coverage or circumstances may still lead to a different
outcome for the current claim.

Models for reserving on individual claims describe how each claim
develops through reporting, payments, recoveries and settlement. \citet{antonio2014micro}
model the timing and amounts of claim development, while
\citet{gabrielli2021individual} combines claim characteristics and payment
histories. \citet{chaoubi2023micro} use a recurrent neural network to predict
whether a payment or recovery occurs in each period and its amount.
Foundation models could enrich these predictors with changes in the text,
such as a new diagnosis or the start of litigation. When a claim is first
reported, the available description could help estimate its ultimate cost;
later notes could update the estimate of remaining payments.
A historical backtest can reconstruct the information available at each
valuation date and compare predictions with and without text. Following the
same claims through later development would show whether text reveals changes
in expected costs earlier and whether the resulting reserve estimates agree
more closely with eventual payments.

\subsection{Vision models}
\label{sec:vision}

Property and vehicle images can reveal physical conditions absent from
structured records. Vision transformers turn images into learned numerical representations
\citep{dosovitskiy2021image}. To identify the part of an image belonging
to a region selected by a user, a model can label the pixels in that
region, a task known as promptable segmentation
\citep{kirillov2023segment}. Vision--language models place images and
text in a shared numerical space so that their representations can be
compared \citep{radford2021learning}. An insurer could use these outputs to identify
physical features for pricing or assess damage for reserving. For example,
an image may show a large tree close enough to fall onto a house during an
extreme storm. An image of an already deteriorating roof may indicate that
hail would cause more damage than on a sound roof, affecting the insurer's
assessment of vulnerability and the need for repairs before offering cover.
Images taken before and after an event could also help distinguish existing
deterioration from new damage.

\citet{blierwong2024images} adapt pretrained image models to obtain
embeddings from street-view imagery and use the embeddings in generalised
linear pricing models. Their study finds improved prediction of claim
frequency for sewer backup and water damage. A related application is early
damage assessment after a storm. Models can classify building damage in
aerial or satellite imagery \citep{ji2018identifying,gupta2019xbd}.
An insurer could link the estimated extent of damage to the buildings it
covers, identify properties needing inspection, and estimate repair costs.
To estimate contractual payments, the insurer can then apply coverage
conditions, deductibles and limits. Combining those estimates across the
affected area could inform reserves and reinsurance recoveries before every
property has been inspected. In this application, the vision model supplies
an early assessment of physical damage; the insurer's cost and coverage
calculations translate that assessment into payments.

\subsection{Geospatial and Earth observation models}
\label{sec:geospatial}

Property rating plans often use postal codes, rating territories or
distance bands, but nearby locations can differ in terrain, vegetation,
drainage and infrastructure. Satellite imagery and geographic data describe
some of these differences at a finer scale. Geospatial foundation models
learn representations from images collected to observe the Earth's surface
\citep{jakubik2023foundation,szwarcman2024prithvi}. An insurer can attach the
resulting embeddings to policy locations and use the embeddings in a loss
model alongside existing geographic variables. The relevant patterns depend
on the peril: slope and vegetation influence wildfire spread, impervious
surfaces and drainage affect water accumulation during heavy rain, and
building layout and tree cover influence storm damage. Representing these
conditions jointly could help distinguish properties that belong to the
same rating territory but face different hazards.

In an earlier ratemaking study, spatial embeddings reduced, in most
settings, both the bias and variance of loss predictions relative to
classical spatial interpolation \citep{blierwong2022geographic}.
Models including SatCLIP \citep{klemmer2023satclip} and AlphaEarth Foundations
\citep{brown2025alphaearth} learn geographic representations that can be
reused beyond one insurer's portfolio. Building on SatCLIP,
\citet{holvoet2025multiview} combine satellite images and OpenStreetMap
features to learn spatial embeddings for risk modelling. To use these spatial embeddings to predict the frequency of flood claims,
an insurer can link them to the properties it covers, their periods of
coverage and the claims observed. After a flood, geographic representations could also
help locate insured buildings in the affected area and prioritise
inspections where flooding is likely to be severe. For agricultural
insurance, representations of land cover and vegetation could contribute to
indices describing growing conditions.

\subsection{Time series and sensor models}
\label{sec:timeseries}

Insurers obtain sensor data from telematics devices and smartphones in
motor insurance, wearables in life and health insurance, household leak and
smoke detectors, and monitors attached to industrial equipment. They also
produce time series of payments, claim counts and exposure. Research on
telematics data relates driving measurements to risk through variables
constructed from speed, acceleration and trip patterns
\citep{verbelen2018unravelling,gao2019claims,gao2022boosting}.
\citet{chan2025telematics} identify anomalies in telematics time series,
while \citet{lee2026telematics} combine the frequency and extremeness of
unusual driving behaviour into a risk index. In the latter study, severity
refers to how unusual the behaviour is, rather than the amount of an insurance
claim.

Time series foundation models, including Chronos \citep{ansari2024chronos}
and TimesFM \citep{das2024decoder}, learn temporal patterns from many series
and reuse those patterns to forecast a new series. Trends, seasonality and
relationships among neighbouring observations may transfer even when the
new series describes a different quantity. TimesFM learns from real and
synthetic series and forecasts from a window of historical observations without updating
its parameters for each task. This approach could shorten the observation
window an insurer needs for a useful forecast and reduce the work required
to construct variables from measurements. Representations of sensor data could support pricing
based on vehicle use or early detection of a leak or impending equipment
failure. Forecasts of future claim payments could help an insurer estimate
the reserves needed for claims already reported.

TimesFM-3 extends this approach to multivariate forecasting, using several
target series, historical covariates and covariates known over the forecast
horizon to produce point and quantile forecasts \citep{jain2026timesfm3}.
In insurance, the target series could be claim counts or payment flows across
product lines and regions, with exposure and calendar information supplied
at the forecast date. Historical insurance data would allow the model to
relate payment patterns in one part of a portfolio to patterns in another.
For capital assessment, an insurer is concerned about unusually large losses
across the portfolio. That application requires modelling how the losses
occur together, so that simultaneous increases across regions or products
enter the estimate of the capital needed to absorb aggregate losses.

A telematics device
may be switched off or lose its connection, leaving gaps in recorded speed
and acceleration. A driver who knows that driving is being monitored may
drive more carefully, so the measurements describe behaviour under the
monitoring programme. Likewise, a leak sensor can prompt a repair before
damage becomes extensive. By acting on model outputs, policyholders can prevent or mitigate losses,
as discussed in Section \ref{sec:economics}. A model trained on policyholders
who voluntarily participate in a monitoring programme may therefore not
transfer to policyholders who choose not to participate. The observation period also limits what can be inferred about future losses. A month of activity data may capture
a temporary routine but miss seasonal behaviour or health changes over a
long contract. By contrast, detecting a leak requires a prompt response to
recent measurements. For these reasons, an insurer should assess how measurements were
collected and whether the observation period matches the insurance
decision. Continuous observation of physical activity and health also
raises questions about privacy and the acceptability of monitoring.

\subsection{Tabular foundation models}
\label{sec:tabular}

Most actuarial data are tabular, but similar columns can describe different
quantities across insurance products. A motor policy may record exposure in
vehicle years, while a property policy covers a building whose insured value
changes over time. Claim amounts may include expenses in one dataset and
exclude expenses in another. A common representation of these variables and
outcomes would help insurers use experience from other portfolios when
developing a new product or modelling a sparsely observed peril.
Pretraining tabular foundation models uses collections of prediction tasks
to learn relationships between observed rows and outcomes for new rows.
Prior-data fitted networks learn from datasets simulated under assumptions
about the processes that could generate the data. The possible processes and their probabilities are described by a prior
distribution. Training across the simulated datasets teaches the
network to approximate predictions that combine the prior with the
observations in a new dataset, known as posterior predictive inference
\citep{muller2022transformers}. TabPFN applies this approach to previously
unseen tables \citep{hollmann2025accurate,zhang2025tabpfn}, and TabICL
classifies observations in large tables using labelled examples supplied
with the prediction request, an approach called in-context classification
\citep{qu2025tabicl}.
Like credibility methods,
these models use information beyond an insurer's small claims sample.
Credibility draws on experience from other comparable insured risks and
combines that collateral information with the insurer's experience through
an updating rule; a pretrained network learns its prediction rule across
training tasks. The analogy concerns borrowing information, rather than a common
weighting formula: a prior-data fitted network uses a prior over processes
that generate data, without an explicit credibility weight.

With only a few observed claims, an insurer can be uncertain about the
parameters of a loss distribution as well as the outcomes of future
claims. Under the Bayesian formulation used by TabPFN, predictions
combine the possible models using probabilities updated by the observed
data. The resulting approximate posterior predictive distribution
combines uncertainty about the fitted distribution with variation in
future outcomes \citep{hollmann2025accurate}. An insurer can use
that distribution to describe possible future claim amounts only insofar as
the synthetic prior represents the insurance task, a limitation discussed
in Section \ref{sec:output-validation}. Assessing how
precisely the conditional mean or a loss quantile has been estimated is a
different calculation. \citet{nagler2026uncertainty} develop an approximate
martingale-posterior method that starts from the PFN predictive distribution
and quantifies uncertainty about quantities such as means and quantiles.
For an insurer entering a new line of business, this distinction separates
the variability of future claims from uncertainty about the average cost or
tail risk of the new portfolio. Nagler and R{\"u}gamer's results concern
independent, identically distributed tabular observations. To account for
dependence, censoring and delayed outcomes in insurance, we propose
pretraining tasks and loss functions that represent those observation
processes, rather than treating every observed payment as a completed,
independent claim.

To predict outcomes for new rows without updating its weights for each
dataset, TabFM uses labelled rows supplied with the prediction request.
It learns this prediction rule from synthetic tables generated by
structural causal models, which specify how variables depend causally
on one another. TabFM combines information across rows and columns
through attention and uses compact representations of rows to perform
classification and regression \citep{kong2026tabfm}. For an insurer with limited claims experience, this offers a
way to use a small labelled portfolio together with relationships learned
during pretraining. A useful comparison would vary the number of available
claims and examine whether TabFM improves estimates relative to models
fitted only to those claims.

To identify the benefit of pretraining tabular foundation models, a
comparison needs both conventional actuarial methods and supervised
learning models with related architectures. Comparing predictions with
those from GLMs and boosted trees can show whether the proposed model
improves on practical alternatives. Attention
models trained on the same insurance portfolio help distinguish the
contribution of broad pretraining from that of attention itself.
\citet{brauer2024transformers} studies transformer models for actuarial
pricing. The Credibility Transformer weights information from the portfolio
together with a policy's covariates in the space of embeddings
\citep{richman2025credibility}. Other supervised learning models provide
different comparisons: tree-like pairwise interaction networks make
interactions interpretable \citep{richman2025treelike}, while Tab-TRM uses
a compact recursive network for claim frequency prediction
\citep{padayachy2026tabtrm}. These models learn from a portfolio of insurance
losses, allowing the value of external pretraining to be examined separately.
In a comparison on French and Belgian motor insurance data,
\citet{deprez2026tabpfn} finds that TabPFN does not consistently improve
frequency or severity prediction over GLMs and XGBoost and requires substantially more computation
to produce predictions in their implementation. Strong general benchmark
results therefore do not yet establish a consistent advantage for
insurance pricing.

A policy observed for three months has had less opportunity to generate a
claim than one observed for a year. A partly paid claim does not yet reveal
its ultimate cost, and an unreported claim is absent from the current
records. Varying periods of exposure, censoring, truncation and delayed
reporting therefore affect what an observed insurance outcome means. A tabular model must account for these differences when
learning from insurance data. For claim counts, exposure can enter through
an offset or another model structure that represents time at risk.
Fine-tuning a pretrained model on insurance claims, changing the synthetic
prediction tasks, and pretraining directly on insurance data offer different
ways to accommodate the observation process. We discuss how to combine
pretrained knowledge with insurer-specific experience, and how to avoid
using future information in that combination, in Section \ref{sec:transfer}.

\subsection{Multimodal models}
\label{sec:multimodal}

A single insurance claim can combine text, images, location data and a
sequence of payments. Separate embeddings represent each source as a numerical vector supplied to a predictive model. A joint embedding could be more informative because it also represents relationships between the sources. For example, an adjuster's description may locate
water damage in a basement while an image shows the extent of the affected
area. A model that processes both sources together could connect the
described location to the visible damage, rather than representing each
source in isolation. Gemini Embedding 2 maps text, images, audio, video and
documents into a shared space and can process combined inputs
\citep{choi2026gemini2}. In a spatial application,
\citet{holvoet2025multiview} combine satellite images with OpenStreetMap
features to learn representations for risk modelling. An actuarial model could use the combined representation
to relate the circumstances of a claim or property to insured losses.

\section{Scientific models and insured losses}
\label{sec:scientific}

Models fitted to insurers' own data primarily describe risks observed in
the past or present. Estimating losses from future weather events and a
changing climate also requires information about how physical conditions
may evolve. Scientific models can simulate weather or the behaviour of a
physical system, supplying scenarios that an insurer can translate into
losses. This is similar to the role of economic scenario generators in
insurance, where simulated interest rates and asset returns feed into valuation
and risk calculations \citep{pedersen2016scenarios}. Physical simulations
instead describe wind, rainfall, temperature or other conditions that affect
insured assets. To estimate insurance payments, the insurer must translate the local
hazard into damage according to the vulnerability of each asset,
then apply repair costs and contract terms to that damage.

\subsection{Weather, climate and digital twins}

A weather forecast can help an insurer anticipate the location and scale
of storm losses before claims are reported. Deterministic weather emulators
approximate numerical weather prediction and return one forecast, such as a
field of wind speeds or precipitation. These models can reduce computation
costs while achieving high accuracy on medium-range forecasting tasks
\citep{bi2023accurate,lam2023learning}. Probabilistic systems instead produce
an ensemble of possible weather outcomes \citep{price2025probabilistic}.
For insurance, the ensemble allows several storm tracks or rainfall patterns
to enter a loss calculation, rather than basing the estimate on a single
forecast. The tail-calibration diagnostics discussed in Section \ref{sec:agenda}
can also assess whether the physical ensemble gives reliable probabilities
for extreme weather at insured locations. Hybrid systems, such as NeuralGCM, combine learned atmospheric
dynamics with physical modelling \citep{kochkov2024neural}. In each case,
the weather model supplies conditions that can be related to the locations
and vulnerabilities of insured assets.

How a model can be reused depends on the tasks and horizons covered by its
training. ClimaX was developed to transfer between weather and climate
tasks \citep{nguyen2023climax}, and Aurora extends weather forecasting to
other Earth system forecasting tasks \citep{bodnar2025foundation}.
Aurora's case studies include forecasts up to ten days, which could inform
preparation for an approaching storm or an early estimate of catastrophe
losses. Decisions about drainage systems or flood defences require a much
longer horizon. NeuralGCM's multidecadal experiments use prescribed
sea-surface temperatures, and its authors report limitations when
extrapolating to substantially different future climates
\citep{kochkov2024neural}. Those limitations constrain the use of its
simulations to value adaptation investments over decades. Lower simulation
costs could nevertheless allow insurers to explore more storm tracks,
rainfall patterns or climate assumptions within a fixed computing budget,
and examine how those alternatives affect insured losses.

Some insurance applications need simulations that reflect the current
condition of a particular asset or region. A digital twin combines
observations of a physical system with a physical or data-driven simulator
\citep{grieves2017digital,rasheed2020digital}. For a floodplain, the
observations might describe river levels, drainage capacity and the state of
local defences. A pretrained model could approximate part of the simulator
or update the simulated state as new measurements arrive. The insurer could
then estimate losses under a rainfall scenario and compare the consequences
of clearing a blocked drain or strengthening a flood defence. This use
requires a model of how the proposed action changes water flow and damage.
World models offer a related approach by learning how an environment changes,
including in response to actions \citep{ha2018world}. Models trained on video
have generated interactive scenes \citep{bruce2024genie}, and learned world models have helped simulated robots and game-playing systems
choose actions by predicting their consequences
\citep{hafner2025mastering}. Accident reconstruction and comparison of
preventive actions are possible insurance extensions. We examine the physical and causal requirements of these insurance
applications in Section \ref{sec:scientific-validation}.

\subsection{From physical conditions to insured losses}

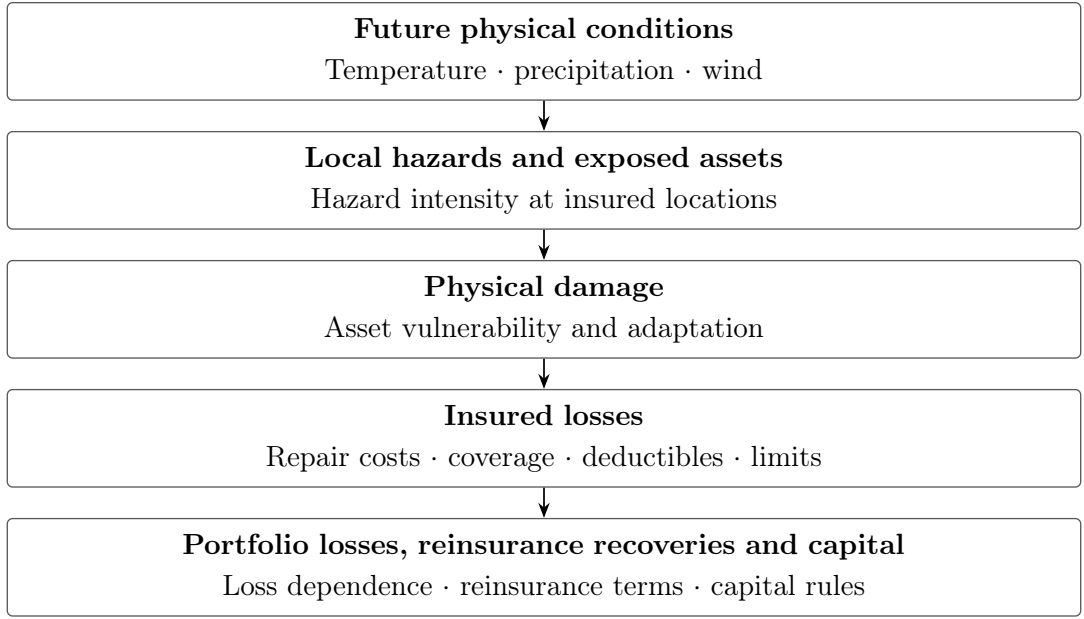
\begin{figure}[!ht]
\centering
\begin{tikzpicture}[
    font=\normalfont\normalsize,
    node distance=4mm,
    box/.style={draw=black!65, line width=0.5pt, rounded corners=2pt,
        align=center, text width=138mm, inner xsep=2mm, inner ysep=2mm},
    arr/.style={-{Stealth[length=2mm]}, line width=0.65pt}
]
\node[box] (conditions)
    {\textbf{Future physical conditions}\\[2pt]
    Temperature \(\cdot\) precipitation \(\cdot\) wind};
\node[box, below=of conditions] (hazards)
    {\textbf{Local hazards and exposed assets}\\[2pt]
    Hazard intensity at insured locations};
\node[box, below=of hazards] (damage)
    {\textbf{Physical damage}\\[2pt]
    Asset vulnerability and adaptation};
\node[box, below=of damage] (losses)
    {\textbf{Insured losses}\\[2pt]
    Repair costs \(\cdot\) coverage \(\cdot\) deductibles \(\cdot\) limits};
\node[box, below=of losses] (portfolio)
    {\textbf{Portfolio losses, reinsurance recoveries and capital}\\[2pt]
    Loss dependence \(\cdot\) reinsurance terms \(\cdot\) capital rules};
\draw[arr] (conditions) -- (hazards);
\draw[arr] (hazards) -- (damage);
\draw[arr] (damage) -- (losses);
\draw[arr] (losses) -- (portfolio);
\end{tikzpicture}
\caption{From physical conditions to insured losses and capital requirements.}
\label{fig:loss-chain}
\end{figure}

To estimate insurance payments from a rainfall forecast, an insurer first
needs information about conditions at each insured property,
such as water depth at the building. A model of local drainage and
terrain can translate rainfall into flooding. To assess the resulting
damage, an insurer can use a vulnerability model, which relates water
depth to damage for each building type. We can then estimate the monetary value of the potential loss as the cost of
repairing that damage. The final payment depends on the policy coverage and the deductible and limits selected by the policyholder. A storm may affect many buildings at once. To estimate losses over the
portfolio, the insurer therefore needs the distribution of losses at
each building and a model of how those losses occur together. A common simulated hazard field already induces dependence between
buildings, so any additional model should describe dependence in damage or
repair costs conditional on that field, rather than count the shared hazard twice. Figure \ref{fig:loss-chain} follows these steps from physical
conditions to portfolio losses, reinsurance recoveries and capital.

Each transformation can change which aspects of the physical forecast
matter for the loss estimate. A small increase in water depth may have little
effect until water reaches an occupied floor, after which damage can rise
sharply. A deductible can absorb small losses, while a policy limit can cap
large payments. Spatial resolution also matters: an average water depth over
a large grid cell may hide the difference between a building on high ground
and one beside a river. The insurer therefore needs a way to translate the
weather model's grid into conditions at insured locations.
Over longer projections, changes in building stock, land use and flood
defences can alter vulnerability. Inflation and demand for labour and materials after a catastrophe affect
repair costs. One can then find the insurer's net loss distribution by combining policy
losses and applying reinsurance terms. For pricing or capital assessment over a specified horizon, the scenarios
also need probability weights and a model of event occurrence and dependence
over that horizon; a short-term forecast ensemble instead supports losses
conditional on the current weather information. The applicable capital framework determines which measure of losses or
changes in the insurer's own funds is used to calculate required capital. This sequence explains
why an accurate physical forecast can still lead to a poor insurance
estimate if vulnerability, costs or contract terms are misrepresented.

\section{Combining foundation and actuarial models}
\label{sec:architecture}

An insurer can use a foundation model's output in a pricing, reserving or risk
calculation. It chooses which information to pass to the actuarial model and
how that model will use it. A language model might supply a litigation indicator to an
existing reserving model, while an image encoder supplies an embedding to a
pricing model. We propose a process that keeps these transformations
explicit, so that an insurer can trace an estimate to the source data and compare
alternative foundation models. We propose a validation framework for this process, including checks on
information use, in Section \ref{sec:governance}.

\subsection{Connecting model outputs to actuarial models}

To connect a foundation model to an actuarial calculation, an insurer
should first choose the output needed for that calculation.
It should define what each named variable represents, such as whether a
litigation indicator means that an action has been filed or merely threatened.
For embeddings, the insurer should select an encoder and any transformation
needed to connect its output to the actuarial model. For retrieval, it
should choose a collection of documents and specify what it wants to find.
For simulation, it should specify assumptions about physical conditions
and the time horizon. The insurer should record these choices
together with the data sources and model versions. It can send a case to a reviewer when the claim description does not support a
reliable value for a required variable. When comparing different embedding spaces, it should instead keep the actuarial
model class, training data and fitting procedure fixed while refitting the
parameters to each representation.

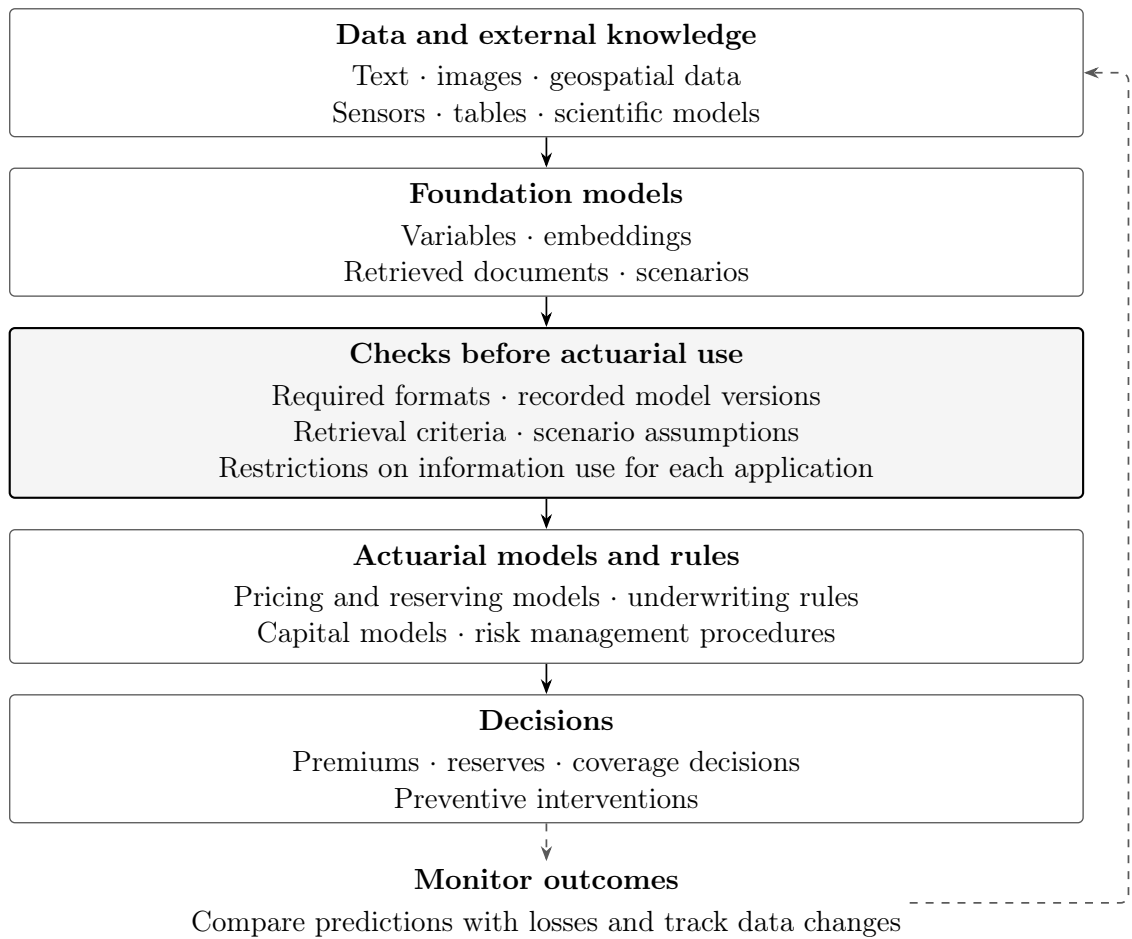
\begin{figure}[!ht]
\centering
\begin{tikzpicture}[
    font=\normalfont\normalsize,
    node distance=4mm,
    box/.style={draw=black!65, line width=0.5pt, rounded corners=2pt,
        align=center, text width=138mm, inner xsep=2mm, inner ysep=2mm},
    arr/.style={-{Stealth[length=2mm]}, line width=0.65pt},
    feedback/.style={arr, dashed, draw=black!65}
]
\node[box] (data)
    {\textbf{Data and external knowledge}\\[2pt]
    Text \(\cdot\) images \(\cdot\) geospatial data\\
    Sensors \(\cdot\) tables \(\cdot\) scientific models};
\node[box, below=of data] (fm)
    {\textbf{Foundation models}\\[2pt]
    Variables \(\cdot\) embeddings\\
    Retrieved documents \(\cdot\) scenarios};
\node[box, below=of fm, draw=black, line width=0.8pt, fill=black!4] (iface)
    {\textbf{Checks before actuarial use}\\[2pt]
    Required formats \(\cdot\) recorded model versions\\
    Retrieval criteria \(\cdot\) scenario assumptions\\
    Restrictions on information use for each application};
\node[box, below=of iface] (act)
    {\textbf{Actuarial models and rules}\\[2pt]
    Pricing and reserving models \(\cdot\) underwriting rules\\
    Capital models \(\cdot\) risk management procedures};
\node[box, below=of act] (dec)
    {\textbf{Decisions}\\[2pt]
    Premiums \(\cdot\) reserves \(\cdot\) coverage decisions\\
    Preventive interventions};
\draw[arr] (data) -- (fm);
\draw[arr] (fm) -- (iface);
\draw[arr] (iface) -- (act);
\draw[arr] (act) -- (dec);
\node[align=center, below=5mm of dec, inner sep=1mm] (outcomes)
    {\textbf{Monitor outcomes}\\[2pt]
    Compare predictions with losses and track data changes};
\draw[feedback] (dec.south) -- (outcomes.north);
\draw[feedback, rounded corners=3pt] (outcomes.east)
    -- (77mm,0 |- outcomes.east) -- (77mm,0 |- data.east) -- (data.east);
\end{tikzpicture}
\caption{From foundation model outputs to actuarial decisions.}
\label{fig:architecture}
\end{figure}

Within our proposed framework, the actuarial model would then use the
prepared information to estimate quantities
such as expected claim costs, remaining payments or a portfolio loss
distribution. The insurer uses those estimates, together with expenses,
capital costs and business constraints, to set premiums, establish reserves
or choose preventive actions. Figure \ref{fig:architecture} shows the steps in this pipeline. A foundation model that directly supplies predictions or predictive
distributions performs part or all of the actuarial prediction step.

To reproduce and compare calculations, an insurer should keep the
encoder's parameters fixed within each deployed version. Fine-tuning or a provider update creates a new version whose outputs may differ. The insurer can recalculate historical cases with the old version and its
settings to identify how much an update changes the result. The insurer's actuaries should oversee the calculation and its use and
monitor predictive performance as new claims are observed.

The information an insurer passes to the actuarial model
also depends on the intended use.
An insurer might use a geographic embedding to assess catastrophe
accumulation but restrict the information allowed into individual pricing.
For each application, the insurer can transform the embedding to select or
modify the inputs supplied to the actuarial model, or direct the case to
review. Such a transformation is a modelling choice, not an assurance that
sensitive information has been removed. We examine both the attributes that can be inferred from the resulting inputs
and how those inputs affect decisions in Section
\ref{sec:information-content}.
The insurer can make the restriction and its consequences visible by recording
the transformation separately for each application.

For a bodily injury claim, suppose a new note says that the claimant has retained counsel. This note can record a change in the claim's legal representation status before any new payment is observed. If that status helps predict remaining payments in past claims, the reserving model can use it to update the estimate before the payment history changes. A language model can turn the note into a variable indicating legal representation and pass that variable to the reserving model together with the claim's other characteristics. To avoid reporting a change in status that the note does not establish, the insurer should distinguish considering legal advice from retaining counsel, and send ambiguous notes for review. It can reproduce the calculation from the dated note and model version.
It can then identify whether a reserve change comes from a different reading
of the same note by replacing the language model while keeping the reserving
model fixed. Even a correctly recorded variable may add little to prediction if existing
claim characteristics already capture the relevant information.

Checking whether a language model has read a claim note correctly
concerns the accuracy of its output. Checking whether that output
improves reserve predictions concerns its use in the actuarial model.
The terminology in \citet{bommasani2021opportunities} distinguishes
intrinsic evaluation of the output itself from extrinsic evaluation of
its contribution to the task. An embedding may improve loss predictions
even though its individual values are difficult to interpret, while a
correctly recorded variable may add no information beyond existing inputs.

Reusing the same encoder for pricing and reserving creates a dependency between the two models. If an error causes that encoder to omit part of the source text, correcting it can change the representations supplied to both models and hence both calculations. The insurer should therefore record which applications use each version
and check how their predictions respond when its outputs change.
We propose comparisons of predictive performance in Section
\ref{sec:output-validation}.

\subsection{Combining pretrained knowledge with insurer experience}
\label{sec:transfer}

A language model can learn how an injury is described, and an image model
can learn patterns associated with buildings. An insurer can then use
its previous claims experience to relate those descriptions and patterns
to payments. Pretraining supplies the representation of the source, while
claims experience supplies the relationship to insured losses.
The insurer can keep the pretrained encoder fixed and fit a loss model to
its outputs, fine-tune some or all of the encoder's parameters on insurance
data, or use a pretrained predictor that accepts a new dataset as context.
These choices change how much the representation adapts to the insurer's
experience. If a portfolio contains too few claims, fitting a large pricing model or fine-tuning an entire encoder risks overfitting. The insurer may still be able to fit a compact pricing model to a fixed encoder's outputs. A larger dataset of claims experience may make more extensive fine-tuning practical.

With few claims, fine-tuning can adapt the encoder to accidental
associations in the insurer's sample and remove patterns that would have
helped on new claims. The aim of learning the insurance application must
therefore be balanced against preserving useful patterns from pretraining. Holding the encoder fixed avoids this particular source of overfitting,
although the actuarial model fitted to its outputs can still overfit. A fixed encoder can also be
poorly suited to the insurer's population: an image model trained mainly on
houses may describe industrial buildings inadequately.
Studies reviewed by \citet{bommasani2021opportunities} examine whether models remain accurate when the data distribution changes. Broad pretraining can improve performance under such changes. For some vision tasks, fitting only the final predictor can give better predictions on a new distribution than adapting the entire model. An insurer can assess how this trade-off changes with sample size and
differences between portfolios by comparing fixed encoders, partial fine-tuning (keeping some of the encoder's weights frozen), full fine-tuning and models trained on its own data.

To reconstruct a reserve estimate at a past valuation date, an insurer
should use only information available at that date. A claim note written
later would reveal something the insurer could not yet have known. Including the note would introduce future information into the comparison between the reserve estimate reconstructed for the valuation date and the payments subsequently observed, a problem known as temporal leakage. Keeping dates with the notes makes
it possible to select the information that was available for each
historical prediction. Filtering claim notes by date does not remove knowledge of later medical,
legal or economic developments learned during pretraining, so a model trained
later may have an advantage over one available at the valuation date.

An embedding computed from existing rating variables can change how a
model uses those variables without adding another source of information. The embedding depends on the original inputs, so it adds no information
conditional on those inputs; it may nevertheless make a useful relationship
easier for a limited prediction model to learn. The same re-encoding can also make a proxy relationship easier to use:
an embedding may help the model exploit a protected attribute already
inferable from the original variables. We discuss how to assess these proxy
relationships and their effects on actuarial predictions in Section
\ref{sec:information-content}. The semantic pricing application of \citet{blierwong2026semantic} illustrates this possibility. The insurer can measure the practical benefit of using the embedding by comparing matched models with and without the embedding.

An insurer can combine its claims experience with information from comparable
risks through credibility methods. An embedding could supply
covariates in a hierarchical loss model, identify a group of similar risks,
or determine parameters of a prior distribution. If the embedding determines
a prior distribution for a risk quantity, the claims can update it through
the likelihood, which describes how the observed claims depend on that
quantity. Using an embedding as a covariate is a different modelling choice.
The fitting procedure must account for reuse of claims data when the same claims help estimate the prior and update the risk quantity. One can estimate the connected model components jointly or use empirical
Bayes methods, in which parameters of the prior are estimated from data.
Treating the two uses of the same claims as independent sources would
overstate the amount of information.

\section{Validation and governance of pretrained models}
\label{sec:governance}

An insurer that adopts a foundation model imports relationships learned
outside its own portfolio. Those relationships can make claim text or
images useful for prediction, but they can also carry biases from the
training data into actuarial calculations. The insurer may have little
control over the data, training objectives or subsequent model updates.
Governance therefore needs to address what the pretrained model brings
into the application: which external relationships transfer, which
personal attributes its outputs retain, and which other applications
share the same source of error. These questions remain even when the
insurer follows established practice for fitting and validating its own
loss model.

\subsection{External training data and transferred bias}
\label{sec:output-validation}

A pretrained encoder learns which features of text or images to represent
before the insurer supplies its claims experience. If its training data
underrepresent a construction type or associate a language variety with
particular social characteristics, the resulting representation may
carry those patterns into pricing or reserving; see
\citet{bommasani2021opportunities} for a broader discussion of how biases
in foundation models pass to downstream applications. Fitting an actuarial model to that
representation does not automatically remove the imported associations.
Their effect depends on what the encoder retains and how the actuarial
model uses its outputs. For example, an image encoder that fails to
distinguish damage on an unfamiliar roof type can deprive the loss model
of relevant information; a text encoder that retains socioeconomic cues
can instead supply information the insurer did not intend to use.

The external assumptions differ across foundation models. Language and
vision models often learn from large collections of observed text and
images. Prior-data fitted tabular models can instead learn from synthetic
prediction tasks generated under a prior
\citep{muller2022transformers,hollmann2025accurate}. For these models, the
question is whether the relationships and observation patterns allowed
by that prior resemble the insurance application. A prior built around
complete, independent rows may transfer poorly to claims observed at
different stages of development or subject to deductibles and limits.
The insurer therefore needs to understand both the populations represented
in observed training data and the assumptions used to generate synthetic
tasks. A large training set alone does not establish that either is
appropriate for its portfolio.

To assess this transfer, we propose comparing a frozen pretrained model,
an adapted version and a model trained on insurer data for the same task.
Holding the insurance evidence and actuarial model comparable helps
identify what external pretraining contributes; varying the number of
claims shows whether that contribution reduces the experience needed.
Results for construction types, languages or regions poorly represented
in pretraining are especially informative about imported errors.
Fine-tuning can change the representation, but a small insurance sample
may supply too little evidence to correct an external association,
especially for rare claims. Improvements in portfolio averages can then
coexist with persistent errors for particular policyholders.

The foundation-model provider's training records may be incomplete or unavailable. The
insurer should record known training populations, periods and objectives,
together with the gaps that prevent it from tracing a suspected bias.
External pretraining also complicates the meaning of an unseen claim:
observations excluded from local fitting may already have appeared in the
provider's training data. Outcomes collected after a documented training
cut-off can supply additional evidence, but cannot retrospectively rule
out overlap in an earlier evaluation. Access to a model's outputs is
therefore insufficient to establish where its predictive knowledge came
from or whether apparent transfer reflects prior exposure.

\subsection{Proxy information in reusable representations}
\label{sec:information-content}

An encoder trained for broad reuse may retain much more than the
information an insurer wants for one task. An image used to assess roof
damage can also show expensive vehicles, while the wording of a claim
note can reveal language background or socioeconomic circumstances.
External pretraining may make these attributes easier to infer from the
resulting embedding. Removing an explicitly recorded protected attribute
from the actuarial dataset therefore does not prevent other inputs from
acting as proxy variables in the final prediction
\citep{prince2020proxy}. The issue is the information carried by the
representation, rather than the names of the fields supplied to the loss
model.

Methods for addressing indirect discrimination provide a starting point
\citep{lindholm2022discrimination,lindholm2024what,frees2023discriminating,cas2022understanding,cote2025fair,charpentier2024insurance}.
Applying them to a reusable embedding requires assessing both the
attributes that can be inferred from it and how those attributes enter
actuarial predictions. To investigate the first question, researchers can
use the embedding coordinates to predict a recorded attribute, such as
language background. Tabular foundation models could make it quick to
compare nonlinear predictions with simpler alternatives
\citep{hollmann2025accurate}. If the tested predictors cannot recover an attribute from the embedding,
that failure does not establish that the attribute is absent: another
model may detect a relationship they missed. Without observations of the
attribute, researchers cannot directly measure how accurately it can be
recovered.

Information can also become usable only when an embedding is combined
with the insurer's structured variables. Let \(Z\) denote the embedding
and \(X\) those variables. Comparing predictions of an attribute from
\(X\), \(Z\) and \((X,Z)\) can reveal whether the embedding adds information
about it, including information that becomes predictive only in
combination with \(X\). For example, location may help interpret a pattern
in a text embedding. Examining the encoder in isolation can therefore
miss a proxy relationship available to the final actuarial model.
Conversely, being able to predict an attribute from the combined inputs
does not establish that it changes premiums or coverage. That conclusion
requires evidence about the actuarial model's use of those inputs;
removing a whole embedding changes many attributes at once and does not
isolate the effect of a single protected characteristic.

Rules on information use may differ between pricing models, reserving
models and models of losses from a common catastrophe. Reusing the same
encoder across these applications can therefore pass information to a
model that is not permitted to use it. We therefore propose recording, for each application,
which pretrained representation enters the calculation, which attributes
can be inferred from it and which restrictions are applied before use.
Transforming an embedding does not by itself demonstrate that a prohibited
attribute can no longer affect the prediction. The relevant comparison
must assess the transformed representation together with the other inputs
and connect changes in decisions to the information the transformation
was intended to remove. This makes governance specific to the combination
of pretrained model, insurer data and actuarial use. When the protected attribute is unobserved, these comparisons cannot
by themselves isolate its effect on decisions; doing so requires additional
data or assumptions about how the attribute relates to the observed inputs.
See \citet{cote2025fair} for a discussion of causal inference and the use
of causal graphs to assess discrimination in insurance pricing.

\subsection{Shared models, prompts and provider changes}
\label{sec:stability}
\label{sec:controls}

A single external model can supply several actuarial applications. If its
provider updates that foundation model and an insurer adopts the new
version, pricing and reserving predictions can change at the same time.
Different commercial services can also depend on the same underlying
model, training data or cloud provider. To identify this dependence, an
insurer needs to connect the applications it uses to their underlying
pretrained components. A fallback supplied by a different vendor may
still share the original model's defect. To continue estimating losses if an embedding service is unavailable or
its encoder misrepresents a construction type, the insurer needs a fallback
actuarial model that does not depend on that encoder, such as a loss model
using only its structured insurance variables. Section \ref{sec:systemic}
examines how widespread use of the same models can also affect the market.

Even without a provider update, a foundation model can respond to how
insurance information is expressed. In semantic pricing,
\citet{blierwong2026semantic} find different predictive performance when
the same policyholder data are expressed through different prompts. Once an application is in
use, a different comparison matters: whether rewording or reformatting
the same information changes the final estimate while the actuarial
model remains fixed. Claim chronology and visible damage must be
preserved in that comparison. The insurer can then measure changes in
reserves or offers of coverage, rather than treating a small numerical
difference between embeddings as evidence of an immaterial effect.
For generative outputs, repeated calculations on identical inputs also
help separate variation from generation itself from sensitivity to the
chosen wording.

When a provider releases a new version of a foundation model, the insurer
needs to compare the old and new versions on the same insurance inputs
before adopting it. For a fixed embedding model, the mapping from inputs
to vectors remains the same. A new version can define a different embedding
space, so an actuarial model fitted to the old vectors may need to be
retrained. The comparison should assess both prediction errors and the
attributes retained by the outputs: an update that improves extraction
accuracy can still change the proxy information available to pricing.

For a proprietary generative model, the provider can change the computation
allocated to reasoning or other settings used to produce responses without
changing the model version presented to customers. Such changes can alter
output quality beyond the random variation inherent in generation;
provider documentation also identifies changes to the model's weights,
infrastructure and configuration as sources of differences between outputs. Earlier tests may therefore overstate the
performance of the service now supplied, and the insurer may be unable to
reproduce an earlier analysis. Access to a fixed version and information
about changes to the service are needed to determine when affected
insurance applications must be reassessed.

\subsection{Transfer of physical simulations}
\label{sec:scientific-validation}

Scientific foundation models import assumptions about physical systems
through their training data and, in some cases, the simulators used to
generate those data. A weather model that performs well on broad forecast
benchmarks may be less accurate in a locality whose terrain or weather
patterns are poorly represented in its training data. Those errors can
lead to incorrect estimates of flooding and insured losses at properties
in that locality. To assess the model's usefulness as a component of a
pricing or loss model, the insurer therefore needs to examine the full
translation from weather to local hazard, damage and covered loss.
Comparisons with numerical or hybrid models can help identify where a
learned forecast fails \citep{rasp2024weatherbench}, but agreement among
models trained on the same simulations is not independent evidence that
the shared assumptions are correct. When the model is used to forecast future climates, weather conditions
may also differ from the historical conditions on which it was trained.

A pretrained simulator used to compare preventive actions must also
represent the physical consequences of the action. For a flood defence,
it must capture how the defence changes water flow and damage, rather
than merely produce a plausible image of the altered landscape. Training
to reproduce observed scenes or trajectories does not by itself establish
this ability to predict interventions \citep{pearl2009causality}. The
insurance assessment must therefore distinguish accuracy on the provider's
original task from evidence that the model supports the proposed change
in the physical system. If a specialised vendor supplies the same physical forecasting model to
many insurers, a local hazard missed by that model may also be missed in
all their loss estimates, even when their actuarial models differ.

\section{Economic and market implications}
\label{sec:economics}

Foundation models can make information already held by insurers usable
for decisions that previously relied on a few structured variables.
Their economic effect depends on what insurers do with the new
distinctions: change prices, recognise liabilities earlier or prevent
damage. Reusing an external model can lower the cost of developing these
applications, while also exposing several insurers to the same errors.
The implications for policyholders therefore extend beyond the predictive
gain measured within one portfolio.

\subsection{Selection, classification and pooling}

A pretrained vision or geospatial encoder can help an insurer distinguish
risks within an existing rating class without developing its own image
model. Suppose it identifies deteriorated roofs that the previous rating
variables did not distinguish. A competitor using that information may
offer lower premiums to owners of sound roofs. If those owners switch,
the original insurer retains a group with higher expected losses than its
common price assumed, which can prompt further repricing. Foundation
models could thus change selection by making such distinctions accessible
to more insurers. The consequences depend on permission to use the
information, competitors' contracts and policyholders' willingness to
switch. This example concerns differences in information between insurers,
whereas adverse selection can also arise when policyholders know more about
their risks than insurers do \citep{akerlof1970market,rothschild1976equilibrium}.

The data available to an encoder can themselves reflect selection.
Lower-risk policyholders disproportionately enter telematics
classification systems in the setting discussed by
\citet{cather2020reconsidering}. A sensor model transferred from those
participants may therefore represent a different population from the
insurer's remaining policyholders. Richer representations can also
support coverage where sparse local experience previously prevented a
useful distinction. A geospatial model might identify effective flood
protection within a territory previously treated as uninsurable. The same
representation could identify other properties for higher prices or
withdrawal; the gain in information does not determine which response
the insurer chooses.

An embedding can make several characteristics, such as roof condition,
construction materials and nearby vegetation, available for pricing at
once. The insurer must decide which of these characteristics to use when
defining risk classes. Pricing
new classes at their conditional expected losses reduces expected
transfers between those classes, while premiums still pool their
unpredictable realised losses.
A household can therefore face a higher premium even though its physical
hazard has not changed: the encoder has made an existing difference in
expected cost usable for classification. The household may lose an
implicit subsidy and find coverage unaffordable
\citep{barry2020personalization,eling2022impact}. The foundation model
expands the set of risk characteristics available for pricing; product
rules, subsidies and the insurer's use of those characteristics determine
how the resulting prices change access to insurance.

\subsection{Prevention, operational value and costs}

The same pretrained representation can support prevention without being
used to create a new rating class. A vision encoder may identify a roof
needing repair, while a time series model may detect a developing leak
from sensor readings. For these applications, the contribution of
pretraining is whether it permits useful warnings with less local
experience or development work. Their value then depends on whether
policyholders act and how much damage their response avoids, after
accounting for false alarms and intervention costs. To choose a preventive action, the insurer needs evidence that acting on
the model's warning, for example by repairing the roof or closing a water
valve, reduces damage. As discussed in Section
\ref{sec:scientific-validation}, identifying existing damage does not by
itself establish the effect of taking action.

Using the same pretrained language encoder for claim retrieval and
reserving can spread the cost of adapting it to insurance records across
both tasks. The
comparison should therefore include the shared model and service costs,
as well as the additional work required for each application. A model
that is cheap to query can still be costly to use if its outputs need
extensive review or if a provider update requires several applications
to be reassessed. The expected value of information offers a way to
compare the resulting decisions with those made without the model
\citep{raiffa1961applied,howard1966information}. In reserving, for example,
the benefit of identifying a worsening injury is earlier recognition of
a liability, rather than a reduction in the underlying injury cost.

\subsection{Dependence on common models}
\label{sec:systemic}

When several insurers use the same foundation model, an error in that
model can enter their calculations together. A shared geospatial encoder
may misrepresent a construction type across their portfolios, or a
language model may mistake threatened legal action for litigation already
filed. The concentration of applications around common pretrained
components is discussed by
\citet[Sections 1.1, 4.7, and 5.6]{bommasani2021opportunities}; the
Financial Stability Board also identifies third-party dependence and
correlated responses as channels of financial risk
\citep{fsb2024artificial}. Different exposures, fine-tuning and actuarial
models can produce different financial consequences from the same error,
so common use alone does not establish identical decisions. Dependence
on foundation models from a few providers nevertheless creates a source
of systemic risk that insurers and supervisors need to address.

Changes to a foundation model can also prompt several insurers to revise
reserves or coverage at the same time. A correction that raises estimated flood risk in one region
could lead several users to increase reserves or reduce coverage there.
If a provider's proprietary foundation model becomes unavailable, insurers
that use it to interpret claim documents may be unable to produce the
outputs needed to process claims, delaying payments at several insurers. These market effects extend the dependence within one insurer
described in Section \ref{sec:controls}. Switching vendors may leave the
exposure unchanged when both services use the same underlying model.
Using several vendors therefore does not diversify the component of model
risk arising from a shared underlying foundation model.

To study the additional dependence created by shared models, researchers
can compare predictions from a shared pretrained component with those
from separately trained alternatives, holding portfolio exposures, storm
scenarios, insurer data and downstream modelling procedures fixed. The comparison would show how sharing the
pretrained component changes the dependence between prediction errors.
Repeating the comparison after each insurer fine-tunes the foundation
model on its own claims would show how much of that dependence remains. Researchers could then calculate how a prediction error common to several
insurers changes the reserves each holds or the coverage each offers,
to assess the financial consequences of relying on the same model. When a provider fixes a defect in a foundation model, insurers that adopt
the corrected version can all benefit from that fix. Using a common model
also spreads development costs across users. The
research question is how much those benefits depend on concentrating
prediction and service provision in a few external models.

\section{Towards actuarial foundation models}
\label{sec:agenda}

A model pretrained on records of different bodily injury claims pooled
from several insurers might require fewer claims from a new portfolio than a model trained only
on that portfolio. Before insurers pool claims, they need to agree on how to protect personal
data and which commercially sensitive information they can share. To assess
whether foundation-model pretraining adds value, we would compare it with
conventional models trained on the same pooled claims. The model would reuse relationships between insurance
inputs and losses, extending the reuse of representations of text or
images. To transfer those relationships to a new insurer's portfolio, the model
would have to distinguish
recurring patterns from differences in coverage, settlement practices
and observation periods. A payment pattern learned elsewhere is useful only
if the model accounts for what that pattern means under the new contract
and claims process. We examine what this broader form of actuarial
prediction could involve and the statistical problems that its development
would need to address.

\subsection{Reusable actuarial prediction}
\label{sec:actuarial-foundation-models}

We use \emph{actuarial foundation model} for a model that learns insurance
relationships across contracts, claim histories or portfolios and adapts
to multiple actuarial tasks. It may predict losses directly or supply
information to another actuarial model. Tabular and time series models
already demonstrate how a predictor can be reused on new data. Prior-data
fitted networks also show how training across simulated datasets can
perform approximate inference under a prior
\citep{muller2022transformers,hollmann2025accurate}. To reuse these predictors across insurance portfolios, the model must
account for how contracts and the process through which claims become
observable change the meaning of the outcome.

For a prediction under an individual contract, the model needs its
coverage, deductible, limit and time horizon. These terms determine how
an underlying loss translates into a covered payment. As a claim
develops, payments already made must be reconciled with the projected
ultimate cost to estimate the amount remaining. At portfolio level, a
model also needs to represent how contracts can incur losses from the
same event. A model might be useful for one of these tasks without
performing all of them. The research objective is to identify which
relationships can be learned across portfolios and reused for a specified
actuarial task.

For individual prediction, let \(Y\) be the insured loss being predicted,
\(X\) the structured information available at the decision date and
\(Z\) an embedding of additional evidence available at that date.
The target \(F_{Y\mid X,Z}\) is the conditional distribution of insured
loss: it describes possible outcomes and their probabilities given those
inputs. An encoder combined with an actuarial model, or a single
pretrained predictor, could estimate that distribution. Since \(Z=h(R)\) is computed from the raw evidence \(R\), conditioning on
\((X,Z)\) can discard information available in \((X,R)\), so the target
distribution also reflects what the encoder fails to retain. The research target is a predictor whose learned relationships remain
useful when these inputs describe a different insurer's contracts and
claims process.

To compare prediction methods, researchers can give competing models
the same information and vary the number of claims available for fitting.
They can then assess when a fixed encoder, fine-tuning or a credibility
method produces useful estimates, and repeat the comparison on the
population to which the model is transferred. This separates the
prediction method from differences in the information supplied to it.

An actuarial foundation model could also learn joint distributions of
losses across contracts. For example, pretraining on geospatial data and
claims from many flood events could help it recognise which properties
are exposed to the same event. When transferred to a new portfolio, the
model would need to combine those learned relationships with the locations
and coverage of the properties now insured to predict losses that occur
together. The research question is which patterns of dependence can be
learned across portfolios and how much local experience is needed to adapt
the resulting joint distribution.

\subsection{Statistical priorities}

An actuarial foundation model must transfer relationships between
insurance records and outcomes across portfolios with different contracts
and claims processes. Below, we identify statistical problems that arise in that
transfer: what uncertainty an external representation carries, which
parts of a predictive distribution remain useful, and how much local
experience is needed to adapt it.

\paragraph{Uncertainty in pretrained representations.}
A pretrained image encoder can return the same vector each time even
when the image does not reveal the full extent of damage. We therefore need to quantify uncertainty in the assessment of that damage. For a deterministic encoder
\(h\), the embedding \(Z=h(R)\) is fixed given the raw input \(R\);
repeating the calculation does not resolve what the image fails to show.
An actuarial model estimating \(F_{Y\mid X,Z}\) must still represent the
remaining variation in loss. If a foundation model instead supplies
a probability distribution over possible damage states, the probabilities
it assigns to those states can enter
the payment calculation through the expected payment for each state. We can use the probabilities of the damage states to calculate the expected
payment by averaging the expected payment for each state. Predicting the full
payment distribution requires combining the conditional payment distributions
across damage states.
The research problem is whether uncertainty learned from external images
remains appropriate for the insured assets and whether dependence among
several model outputs is preserved.

\paragraph{Calibration after transfer.}
A predictive distribution learned on one portfolio may be miscalibrated
on another because deductibles, limits and settlement practices change
what is observed. In particular, the probability of zero payment or a
payment at the policy limit can change even when the underlying damage
model remains useful. An actuarial foundation model needs to separate
reusable loss relationships from these changes in policy terms. To transfer predictions across policy modifications, we need to account
for how those modifications change payments. If the recorded payments are
censored or truncated, we may need additional data or explicit modelling
assumptions to estimate the loss distribution under the modified policy. Conformal
methods can adjust a pretrained predictor using local calibration data,
but their coverage guarantees still require the relevant exchangeability
conditions \citep{angelopoulos2023conformal}. The relevant exchangeability is between calibration and future claims
within the target portfolio; temporal change, delayed outcomes and selection
of settled claims can violate this condition even after external pretraining. Embeddings could help define groups
for local calibration, with the grouping rule fixed independently of
the calibration sample.
We propose comparing how many local claims each pretrained representation
requires to obtain narrow prediction intervals that attain the chosen
coverage level in the new portfolio. This comparison should also examine
whether coverage deteriorates for groups with few claims or claims whose
final costs are not yet observed. This comparison assesses prediction-interval coverage, which does not
establish calibration of the full conditional distribution or its tails;
the latter require the separate diagnostics discussed below.

\paragraph{Learning about large losses.}
Large losses are rare in an insurer's own records, so a central question
is whether pretraining can reduce the number of such claims needed to
predict them. An embedding may extract more information from an existing
claim description, but it does not supply another observed large loss. External insurance
records can add such observations; physical or synthetic simulations add
information under their generating assumptions. An actuarial foundation
model must distinguish those sources when transferring a relationship
to a portfolio with little experience of large losses. Embeddings could
supply covariates for extreme-value regression.
Comparing representations at a fixed number of extreme claims would
identify whether pretraining makes those claims more informative for
prediction. Varying that number would then show how much local experience
is needed, rather than attributing all gains to a larger training sample.
After transfer, we also need to assess whether the pretrained model's
forecasts remain reliable for extreme losses. The tail-calibration
diagnostics of \citet{allen2025tail} could help evaluate both the predicted
frequency of losses above a high threshold and the predicted distribution
of their sizes beyond that threshold.

\paragraph{Updating the representation or the actuarial model.}
When inflation raises repair costs, an encoder may still correctly
represent physical damage while the actuarial model needs to be updated
to translate that damage into higher payments. If an insurer begins covering construction types absent from the encoder's
training data, it may instead need to fine-tune the encoder on images of
those structures.
A change in how adjusters describe losses in their claim notes can alter
text embeddings even when
underlying risks remain similar. These cases call for different updates:
recalibrating the actuarial model, adapting the representation or changing
how documents are prepared. Researchers can compare those choices by
the new claims experience and computation needed to restore predictive
performance. Keeping the external model version identifiable is essential
to distinguish a provider's change from a change in the portfolio.
The aim is to learn which relationships survive transfer and which parts
of the application need to be trained again.

\subsection{Data and an initial reserving application}
\label{sec:stakeholders}

To compare transfer across insurers, researchers need datasets that link
source documents and measurements to contract terms, observation dates
and eventual payments. Public datasets used for pretraining or evaluating
encoders often supply only part of that chain. For example, xBD contains
satellite images, building footprints and damage labels
\citep{gupta2019xbd}, while an insurance application also needs coverage
terms, repair costs and claim development. Linking these records across
insurers would allow researchers to test whether a representation remains
useful under different contracts and claims processes.

One potential initial application is reserving for bodily injury claims:
pretrained
text representations could identify deterioration before it appears in
payment history. To measure the value of text beyond the adjuster's recorded case estimate,
we would compare a reserving model using structured records that include
that estimate, one also using text representations learned from the insurer's
own claims, and one using a pretrained text model. Varying the
number of local claims would show whether external pretraining helps
recognise changes in remaining payments with less insurance experience.
This separates the benefit of reading the notes from the benefit of
having learned language relationships elsewhere. Fit and tune only on information available by each valuation date, keep
all snapshots of a claim in one data split, and account for incomplete and
unequal follow-up when evaluating unsettled claims. Testing today's pretrained
model on historical records assesses retrospective performance, whereas
reconstructing a system deployable at the historical date also requires
restricting pretraining to information then available, as discussed in
Section \ref{sec:transfer}.
\citet{baillargeon2021mining} use recurrent neural networks to extract
predictors from accident descriptions for modelling the number of vehicles
involved. An actuarial foundation model could extend this use of claim
text to reserving by learning from larger collections of insurance records
and using more capable pretrained language models.
We could also test a predictor trained on one insurer's portfolio on
another insurer's portfolio, keeping the encoder fixed or fine-tuning it
with increasing amounts of local experience. Differences in treatment descriptions, settlement
practices and contract terms would test which learned relationships remain
useful.

\section{Conclusion}
\label{sec:conclusion}

Foundation models could help insurers use details that are already
present in claim descriptions, images and measurements but difficult to
represent through a small set of codes. An embedding can carry those
details into a loss model, while extracted variables and retrieved
documents support calculations and human judgments in more explicit
forms. Simulations from physical models extend the available information to future
weather and climate conditions. To use a simulated event in an insurance
calculation, an insurer must connect the local hazard to physical damage,
repair costs and the payments covered by the contract. Existing studies
show benefits in some insurance applications. Broader use of actuarial
predictors across portfolios would require the learned relationships
to remain useful despite differences in contracts, populations and the
way claims are observed.

We propose a process for using foundation model outputs to predict
insured losses and support premium setting under insurance contracts, with
the source data, transformations and model versions specified for each application. The additional governance burden comes from importing relationships
learned on external data, allowing proxy information to pass through
reusable representations and relying on components shared across
applications and insurers. Local adaptation and predictive accuracy do
not by themselves establish that these inherited relationships are
appropriate for every insurance use. Actuaries therefore need to decide how richer data should enter insurance
calculations, as well as fit the predictor, complementing the vision in \citet{richman2024vision}.

The potential benefits come through earlier recognition of changing
claim costs, more detailed assessment of physical risks and better
targeted prevention. Their value depends on the decisions made with
the information and on policyholders' responses. Finer classification
can change expected transfers and access to coverage, while reliance
on common providers can expose several insurers to the same error or
interruption. Developing actuarial foundation models therefore requires
both statistical methods suited to insurance data and an account of how
the resulting predictions change decisions.

% Bibliography embedded from main.bbl for single-file submission.

\end{document}